\documentclass[prd,onecolumn,nofootinbib]{revtex4}
\usepackage{bm,amsmath,amssymb,graphicx}

\begin{document}

\title{Gravitational collapse in the expanding Universe}
\author{Nikodem Pop{\l}awski}
\affiliation{Department of Mathematics and Physics, University of New Haven, West Haven, CT, USA}
\altaffiliation{NPoplawski@newhaven.edu}

\begin{abstract}
We use the Tolman metric to describe gravitational collapse of a sphere of a fluid without pressure in spacetime with the Hubble parameter $H$ related to the cosmological constant.
We show that the largest radius of a galaxy formed from such a fluid with mass $M$ is given by $(GM/H^2)^{1/3}$.
\end{abstract}
\maketitle

\noindent
{\bf Tolman metric}.\\
The gravitational field of a centrally symmetric collapsing body is given by the Tolman metric \cite{Tolman,LL2}:
\begin{equation}
    ds^2=e^{\nu(\tau,R)}d\tau^2-e^{\lambda(\tau,R)}dR^2-e^{\mu(\tau,R)}(d\theta^2+\mbox{sin}^2\theta\,d\phi^2),
    \label{grav1}
\end{equation}
where $\nu$, $\lambda$, and $\mu$ are functions of a time coordinate $\tau$ and a radial coordinate $R$, and $c=1$.
Applying coordinate transformations $\tau\rightarrow \tilde{\tau}(\tau)$ and $R\rightarrow \tilde{R}(R)$ does not change the form of the metric (\ref{grav1}).
The components of the Einstein tensor corresponding to (\ref{grav1}) that do not vanish identically are \cite{Tolman,LL2}:
\begin{eqnarray}
    & & G_0^0=-e^{-\lambda}\Bigl(\mu''+\frac{3}{4}\mu'^2-\frac{1}{2}\mu'\lambda'\Bigr)+\frac{1}{2}e^{-\nu}\Bigl(\dot{\lambda}\dot{\mu}+\frac{1}{2}\dot{\mu}^2\Bigr)+e^{-\mu}, \nonumber \\
    & & G_1^1=-\frac{1}{2}e^{-\lambda}\Bigl(\frac{1}{2}\mu'^2+\mu'\nu'\Bigr)+e^{-\nu}\Bigl(\ddot{\mu}-\frac{1}{2}\dot{\mu}\dot{\nu}+\frac{3}{4}\dot{\mu}^2\Bigr)+e^{-\mu}, \nonumber \\
    & & G_2^2=G_3^3=-\frac{1}{4}e^{-\nu}(\dot{\lambda}\dot{\nu}+\dot{\mu}\dot{\nu}-\dot{\lambda}\dot{\mu}-2\ddot{\lambda}-\dot{\lambda}^2-2\ddot{\mu}-\dot{\mu}^2) \nonumber \\
    & & -\frac{1}{4}e^{-\lambda}(2\nu''+\nu'^2+2\mu''+\mu'^2-\mu'\lambda'-\nu'\lambda'+\mu'\nu'), \nonumber \\
    & & G_0^1=\frac{1}{2}e^{-\lambda}(2\dot{\mu}'+\dot{\mu}\mu'-\dot{\lambda}\mu'-\dot{\mu}\nu'),
    \label{grav2}
\end{eqnarray}
where the dot denotes partial differentiation with respect to $\tau$ and the prime denotes partial differentiation with respect to $R$.

In a comoving frame of reference, the spatial components of the four-velocity vanish.
We consider gravitational collapse in spacetime with a cosmological constant $\Lambda$.
If the matter composing the body is an ideal fluid with energy density $\epsilon$ and pressure $p$, then the components of the energy--momentum tensor that do not vanish are: $T^0_0=\epsilon+\Lambda/\kappa$, $T^1_1=T^2_2=T^3_3=-p+\Lambda/\kappa$, where $\kappa=8\pi G$.
The Einstein field equations $G^i_k=\kappa T^i_k$ in this frame of reference are:
\begin{equation}
    G_0^0=\kappa\epsilon+\Lambda,\quad G_1^1=G_2^2=G_3^3=-\kappa p+\Lambda,\quad G_0^1=0.
    \label{grav3}
\end{equation}
The covariant conservation of the energy--momentum tensor gives
\begin{equation}
    \dot{\lambda}+2\dot{\mu}=-\frac{2\dot{\epsilon}}{\epsilon+p},\quad \nu'=-\frac{2p'}{\epsilon+p},
    \label{grav4}
\end{equation}
where the constants of integration depend on the allowed transformations $\tau\rightarrow \tilde{\tau}(\tau)$ and $R\rightarrow \tilde{R}(R)$.\\

\noindent
{\bf Dustlike sphere with a cosmological constant}.\\
If the pressure is homogeneous (no pressure gradients), then $p'=0$ and $p=p(\tau)$.
This condition is satisfied for a nonrelativistic fluid (dust): $p=0$.
In this case, the second equation in (\ref{grav4}) gives $\nu'=0$.
Therefore, $\nu=\nu(\tau)$ and a transformation $\tau\rightarrow \tilde{\tau}(\tau)$ can bring $\nu$ to zero and the metric tensor component $g_{00}=e^\nu$ to 1.
The coordinate system becomes both comoving and synchronous \cite{LL2}.
Defining $r(\tau,R)=e^{\mu/2}$ turns (\ref{grav1}) into
\begin{equation}
    ds^2=d\tau^2-e^{\lambda(\tau,R)}dR^2-r^2(\tau,R)(d\theta^2+\mbox{sin}^2\theta\,d\phi^2).
    \label{grav5}
\end{equation}
The Einstein equations (\ref{grav2}) reduce to
\begin{eqnarray}
    & & \kappa\epsilon+\Lambda=-\frac{e^{-\lambda}}{r^2}(2rr''+r'^2-rr'\lambda')+\frac{1}{r^2}(r\dot{r}\dot{\lambda}+\dot{r}^2+1), \nonumber \\
    & & \Lambda=\frac{1}{r^2}(-e^{-\lambda}r'^2+2r\ddot{r}+\dot{r}^2+1), \nonumber \\
    & & 2\Lambda=-\frac{e^{-\lambda}}{r}(2r''-r'\lambda')+\frac{\dot{r}\dot{\lambda}}{r}+\ddot{\lambda}+\frac{1}{2}\dot{\lambda}^2+\frac{2\ddot{r}}{r}, \nonumber \\
    & & e^{-\lambda}(2\dot{r}'-\dot{\lambda}r')=0.
    \label{grav6}
\end{eqnarray}

Integrating the last equation in (\ref{grav6}) (with an integrating factor $r'$) gives
\begin{equation}
    e^\lambda=\frac{r'^2}{1+f(R)},
    \label{grav7}
\end{equation}
where $f$ is a function of $R$ satisfying a condition $1+f>0$ \cite{LL2}.
Substituting (\ref{grav7}) into the second equation in (\ref{grav6}) gives $2r\ddot{r}+\dot{r}^2-f=\Lambda r^2$, which is integrated (with an integrating factor $\dot{r}$) to 
\begin{equation}
    \dot{r}^2=f(R)+\frac{F(R)}{r}+\frac{1}{3}\Lambda r^2,
    \label{grav8}
\end{equation}
where $F$ is a positive function of $R$.
Substituting (\ref{grav7}) into the third equation in (\ref{grav6}) does not give a new relation.
Substituting (\ref{grav7}) into the first equation in (\ref{grav6}) and using (\ref{grav8}) gives
\begin{equation}
    \kappa\epsilon=\frac{F'(R)}{r^2 r'}.
    \label{grav9}
\end{equation}
Combining (\ref{grav8}) and (\ref{grav9}) gives
\begin{equation}
    \dot{r}^2=f(R)+\frac{\kappa}{r}\int_0^R\epsilon r^2 r'dR+\frac{1}{3}\Lambda r^2,
    \label{grav10}
\end{equation}
which determines the function $r(\tau,R)$ \cite{collapse}.
Integrating this equation gives another function, $\tau_0(R)$.
Gravitational collapse of a dustlike sphere in spacetime with a cosmological constant therefore depends on three arbitrary functions of $R$: $f$, $F$, and $\tau_0$ \cite{LL2,condensation}.\\

\noindent
{\bf Gravitational collapse}.\\
Every particle in a collapsing fluid sphere is represented by a radial coordinate $R$ that ranges from 0 (at the center of the sphere) to $R_0$ (at the surface of the sphere).
The function $r(\tau,R)$ is the distance of a particle with coordinate $R$ from the center.
If the mass of the sphere is $M$, then the Schwarzschild radius $r_g=2GM$ of the black hole that forms from the sphere is equal to \cite{LL2}
\begin{equation}
    r_g=\kappa\int_0^{R_0}\epsilon r^2 r'dR.
    \label{grav11}
\end{equation}
Substituting (\ref{grav9}) into (\ref{grav11}) gives
\begin{equation}
    r_g=F(R_0)-F(0)=F(R_0),
    \label{grav12}
\end{equation}
which determines the value of $R_0$ if the function $F$ is given.

If $b(\tau)$ is the radius of the sphere (the distance of a particle with coordinate $R_0$ from the center):
\begin{equation}
    b(\tau)=r(\tau,R_0),
    \label{grav13}
\end{equation}
then equations (\ref{grav10}) and (\ref{grav11}) give
\begin{equation}
    \dot{b}^2(\tau)=f(R_0)+\frac{r_g}{b(\tau)}+\frac{1}{3}\Lambda b^2(\tau).
    \label{grav14}
\end{equation}
If $b_0=b(0)$ is the initial radius (\ref{grav13}) of the sphere  and the sphere is initially at rest, then $\dot{b}(0)=0$.
Consequently, the relation (\ref{grav14}) determines the value of $R_0$ if the function $f$ is given \cite{collapse}:
\begin{equation}
    f(R_0)=-\frac{r_g}{b_0}-\frac{1}{3}\Lambda b_0^2.
    \label{grav15}
\end{equation}

\noindent
{\bf Largest size of a collapsing sphere}.\\
The dynamics of gravitational collapse of a dustlike sphere in the presence of a cosmological constant is determined by equation (\ref{grav8}).
That equation can be solved analytically for $\Lambda=0$ \cite{Tolman,LL2,collapse}, giving also an approximated solution for $\Lambda>0$ if $b_0\ll(r_g/\Lambda)^{1/3}$.

If $b_0$ is sufficiently large, on the order of magnitude of $(r_g/\Lambda)^{1/3}$, the outermost parts of the sphere will expand instead of contracting because of the expansion of the Universe.
Substituting (\ref{grav15}) into (\ref{grav14}) gives
\begin{equation}
    \dot{b}^2(\tau)=\frac{r_g}{b(\tau)}+\frac{1}{3}\Lambda b^2(\tau)-\frac{r_g}{b_0}-\frac{1}{3}\Lambda b_0^2.
    \label{grav16}
\end{equation}
For small $\tau$ (near the initial time $\tau=0$), putting
\begin{equation}
    b(\tau)=b_0+\delta b(\tau),\quad \delta b\ll b_0
    \label{grav17}
\end{equation}
into (\ref{grav16}) and omitting terms of higher order in a small quantity $\delta b$ gives
\begin{equation}
    \dot{b}^2=\Bigl(\frac{2}{3}\Lambda b_0-\frac{r_g}{b_0^2}\Bigr)\delta b.
    \label{grav18}
\end{equation}
The entire sphere will collapse, $\delta b<0$, if $2\Lambda b_0/3<r_g/b_0^2$, which gives
\begin{equation}
    b_0\le\Bigl(\frac{3r_g}{2\Lambda}\Bigr)^{1/3}.
    \label{grav19}
\end{equation}

\noindent
{\bf Largest size of a galaxy}.\\
The observed Universe is dominated by dark energy, which is most naturally described by a positive cosmological constant.
The current value of the Hubble parameter $H$ is approximately related to the cosmological constant by $\Lambda=3H^2$ (asymptotically, $H$ tends to $(\Lambda/3)^{1/2}$).
The relation (\ref{grav19}) can thus be written as
\begin{equation}
    b_0\le\Bigl(\frac{GM}{H^2}\Bigr)^{1/3}.
    \label{grav20}
\end{equation}
This relation determines the largest radius of a spherical galaxy (or approximately other galaxies) because even if not all the matter forms a central black hole and most of it forms stars which are in orbital motion around the galaxy center, $b_0$ is related to $r_g$ corresponding to the entire mass of the galaxy (the gravitational field at radius $b_0$ is determined by the entire mass within this radius).

The observed Hubble parameter is $H\sim 10^{-18}$ s$^{-1}$ (corresponding to the cosmological constant $\Lambda\sim 10^{-52}$ m$^{-2}$) \cite{obs}.
For the Milky Way, the mass is $M\sim 10^{12}$ solar masses and the radius is $b\sim 10^{21}$ m.
According to (\ref{grav20}), the radius should be closer to $b\sim 10^{22}$ m.
However, this value determines an upper limit on the size of a galaxy in the Universe in the presence of a positive cosmological constant, accelerating with a constant Hubble parameter.
The current Universe has not yet reached exponential expansion, driven by the cosmological constant alone.
Therefore, the upper limit on the size of the Milky Way is one order of magnitude higher than its actual size.

The results of this work are also valid in the Einstein--Cartan theory of gravity \cite{EC}, which removes the symmetry constraint on the affine connection and relates its antisymmetric part, the torsion tensor \cite{Niko}, to the intrinsic angular momentum (spin) of the matter composed of fermions, described by the Dirac equation.
In the presence of torsion, which manifests at extremely high densities as a repulsive force, gravitational singularities in black holes may be eliminated, and the singular big bang may be replaced with a regular big bounce \cite{universe}.
In vacuum, torsion vanishes and this theory reduces to general relativity, passing all its observational tests.
Furthermore, torsion violates the commutativity of translation, so the momentum operator components do not commute in spacetime with torsion.
Divergent momentum integrals in Feynman diagrams are thus replaced with convergent sums, removing ultraviolet divergence in quantum electrodynamics \cite{regular}.

I am grateful to Francisco Guedes and my Parents, Bo\.{z}enna Pop{\l}awska and Janusz Pop{\l}awski, for supporting this work.

\end{document}